Kids Today: Remote Education in the time of COVID-19

Adriana Mejia Castaño, Javier E Hernandez, Angie Mendez Llanos

ABSTRACT. With the recent COVID-19 breakup, it became necessary to implement remote classes in schools and universities to safeguard health and life. However, many students (teachers and parents, also) face great difficulties accessing and staying in class due to technology limitations, affecting their education. Using several nationally representative datasets in Colombia, this article documents how the academic performance of students in their final high school year is affected due to technologies, aggregated by municipalities. We conclude that internet access strongly affects these results, and little improvement on the internet/computer access will reflect better academic performance. Under these conditions, belonging to an ethnic group or high rurality (non-geographic centralized municipalities) has a negative impact. Policy implications are discussed.

INTRODUCTORY PARAGRAPH. Governments across the world ordered measures for management, control, and prevention of Covid-19, especially how to proceed for a remote academic continuity program. UNESCO on its website has several resources for distance learning solutions, as resources to provide psychosocial support, digital learning management systems, Massive Open Online Course (MOOC) Platforms and self-directed learning content, and so on (UNESCO, 2020). According to Hartley and Bendixen (2010), there is strong evidence that learning has become more accessible due to the internet and helps students by improving their academic achievement (Sazili & Khafidhah, 2017). However, implementation of these policies or platforms requires (Zhao, Lu, Huang & Wang, 2010):

- Learning skills, self-regulatory skills, epistemological beliefs, motivation, self-efficacy, ability, physical challenges, and learning disabilities.
- The autonomy of use, since the use of the internet is not properly controlled, could have a negative impact (Sazili & Khafidhah, 2017).
- Social and parental support.
- Hypermedia learning material as videos, music, images, and e-books.
- Access and quality of access to the internet, computer, mobile telephone, or tablet for each student.

In particular, these implementations could increment the gap in access to quality education in low-income or geographically remote populations, which keeps countries from achieving sustainable development goals from the United Nation Development Programme (Zhao, Lu, Huang & Wang, 2010).

Despite the need to establish these remote programs, there is little empirical evidence on how technologies in the last years affect their performance (Zhao, Lu, Huang & Wang, 2010) and what is the effect of its absence. This is a critical question given that in pandemic times almost all academic activities require internet access and all economic activities resulted affected (presented reductions) due to the pandemic (Chetty, Friedman, Hendren & Stepner, 2020).

In Colombia, data sets like those provided by the ICFES (Colombian Institute for the Evaluation of Education) have long comparisons over time about the academic achievement of students in all stages. Most of the studies over these data indicate that the covariables that most affect the academic performance (have a good score), according to Chica, Galvis, and Ramirez (2011) are socioeconomic status, parents scholarship, the number of hours in the school, school type (private or public) and gender; but there is little work on how technologies affect it.

The current article aims to fill this gap. Using nationally representative samples of children in their final high school year from 2014 to 2019, this article addresses two related research questions:

- Which municipalities have the most difficulty in implementing the remote academic continuity program?
- What should be the municipalities' prioritization (in each state) for information technology programs investment?

We decide to proceed using three classification algorithms and compare them to predict academic performance. Solving these questions will help decision-makers focus on certain geographical areas to develop specific intervention programs to improve the learning experience and results of high school and middle school students.

BACKGROUND. In this era of high internet connectivity, which brings positive and negative impacts in our lives, many facets are linked to its access; and education is not an exception. Since the early 2000s, it has been noticed (Pereira, Pleguezuelos, Meri, 2007; Greenhow, Robelia, Hughes, 2009) that there were necessary changes to take advantage of the online resources. Parents and teachers know the importance and necessity of computer usage and internet access in all stages of academic formation (Bassok, Latham, 2017).

In kindergarten students in the United States, Bassok, and Latham (2017), found that computer access increases academic children skills. The results of Amorim, Jeon, Abel, Felisberto, Barbosa, and Martins (2020), in Brazil, show that technologies, specifically using games, improves reading scores.

In high school students, specifically in China, the relation between internet self-efficacy, internet accessibility, behavior, academic performance between others was investigated (Zhao, Lu, Huang & Wang, 2010; Penaranda, Aragon-Muriel, Micolta, 2014). There is strong evidence that digital inequality exists on the internet, and students with internet access at home had a better academic performance. In the United States, Rickles, Heppen, Allensworth, Sorensen, and Walters (2018), found no statistically significant differences in longer-term outcomes between students taking online and face-to-face courses; this could indicate that online courses have the same benefits in some populations (a majority of Hispanic and African American students).

In college students, Junco and Cotten (2012), and Sazili and Khafidhah (2017) found a negative correlation between academic performance and internet usage, and a weak relationship with computer ownership.

In university students, Sushma, Draus, Goreva, Leone, and Caputo (2014) pointed out that time spent on the internet could be a measure of academic achievement under certain conditions, however, some online media could have a negative effect on university students (Junco, 2012; Asemah, Okpanachi &

Edegoh, 2013). According to Paulsen and Mccormick (2020), student engagement is necessary to have positive benefits. In countries like Malaysia and Taiwan, the use of the internet is apparently a distractor. Indeed, the results of Xu, Wang, Peng, and Wu (2019) indicate that discipline plays a vital role in the academic success of undergraduate students; in particular, the results show that the use of Internet data could differentiate and predict student academic performance.

Therefore, the evidence suggests that in any stage of learning, technology strongly affects academic performance across the world (Donald, Foehr, 2008) and it is most notorious in pandemic times since all economic activities were seriously affected (Chetty, Friedman, Hendren & Stepner, 2020) and internet access became a necessity (Chiou and Tucker, 2020). Going deeper, it looks like there is a way to predict academic performance based on some covariables (Hellas et.al, 2018), and technologies could be one of them.

In Colombia there is an academic test provided by ICFES, proctors on standardized tests, called SABER11, that scores students in their final high school year, and also has self-reported socio-demographic information. Analyzing data from 2009, Chica, Galvis, and Ramirez (2011) found that high school graduates who own a computer are more likely to obtain a high grade on SABER11, although internet possession was not statistically significant. However, analyzing more recent data, internet access became an interest covariable (see Appendix). In particular, family-related variables are the best to predict academic performance (Garcia-Gonzalez, Skryta, 2019). The dataset from ICFES also includes information of undergraduate students, and an analysis of academic efficiency about these students shows that social covariables play an important role (Rojas 2019; Delahoz-Dominguez, Zuluaga, Fontalvo-Herrera, 2020).

During pandemic times it is necessary to understand how the internet or computer access affects the score of SABER11 since that can offer a way to comprehend the effect of these technologies in each municipality and which are the best improvements according to each necessity. The actual existing evidence, while limited, does support the notion that internet/computer access strongly affects the academic performance of SABER11. The findings suggest that, as we notice in literature, technologies strongly affect academic performance. However, in this scenario, there are two covariables that need to be in consideration: rurality and belonging to an ethnic group. Then upgrades in technology could improve academic performance, but it is not enough: teachers, parents, and the social environment play an important role. There are necessary policies about technologies and, like Zhao, Lu, Huang, and Wang (2010) and Sazili and Khafidhah (2017) pointed out, about increasing learning skills in remote environments.

METHOD.

Data.

SABER11 tests have been modified during the last 20 years, we have chosen the period from 2014 onwards since it has the same standardized content, taking tests on 5 different subjects: critical reading, citizenship skill, English, written communication, and quantitative reasoning; score from 0 to 100. Their sum gives a global score from 0 to 500 (ICFES, 2020). This is the independent variable that we consider in this study. This data also has socio-demographic information such as internet accessibility, computer, family income, habitability conditions, belonging to an ethnic group, among others. There are 8 municipalities that in this period did not take the SABER11 test, which are not included in this work. This data was aggregated by municipality into numerical covariables.

Given that data about internet ownership is self-reported by students, we also decided to use the data reported by the Ministry of information technologies and communications MINTIC, about the total number of internet subscribers, the type of subscription and technology, indicators of internet reach by region, information on internet providers by demand and participation statistics (MINTIC, 2020).

Additionally, there is relevant information by region reported by the DANE (National Administrative Department of Statistics), including accurate population values (2018 Census). With this information it is possible to know how rural a municipality is, to compare if this rurality is correlated with scores, internet access, and other important covariables (DANE, 2020).

In the Appendix, we collect descriptive statistics from a consolidated data frame aggregate by municipalities. From this exploratory data analysis and since we need covariables that allows governmental investments, we conclude that the main numerical covariables to use in this study are:

- CODE: numeric code that identifies the municipality,
- YEAR: year of the proctored test,
- INTERNET: percentage of students with self-reported internet,
- COMPUTER: percentage of students with self-reported computers,
- ETHNIC: percentage of students self-reported as belonging to an ethnic group,
- SCHOOL: percentage of public schools,
- GLOBAL_SCORE: student global score average on SABER11,
- POPULATION: censused population during the test year,
- CONNECTIVITY: access to the internet per 1000 inhabitants,
- RURAL_INDEX: percentage of the population that lives in rural sectors.

Analysis.

A first finding of the exploratory data analysis, see Appendix, is that if the family owns a computer, the average score is higher but RURAL INDEX decreases, coinciding with the findings of Chica, Galvis, and Ramirez (2011). The results showed that scores have a mainly rural composition. Comparing this with geographical location (Figure 1), we obtain that good scores are geographically centralized, there are many families without the internet, and it looks like these two variables are correlated.

In this kind of study, the variables are commonly analyzed via correlation (Sazili & Khafidhah, 2017; Sushma et.al., 2014). In Figure 2, we can see a heat map with the correlations of the initial set of covariables. We see a positive correlation between INTERNET, COMPUTER, and CONNECTIVITY. ETHNIC has a negative correlation with GLOBAL

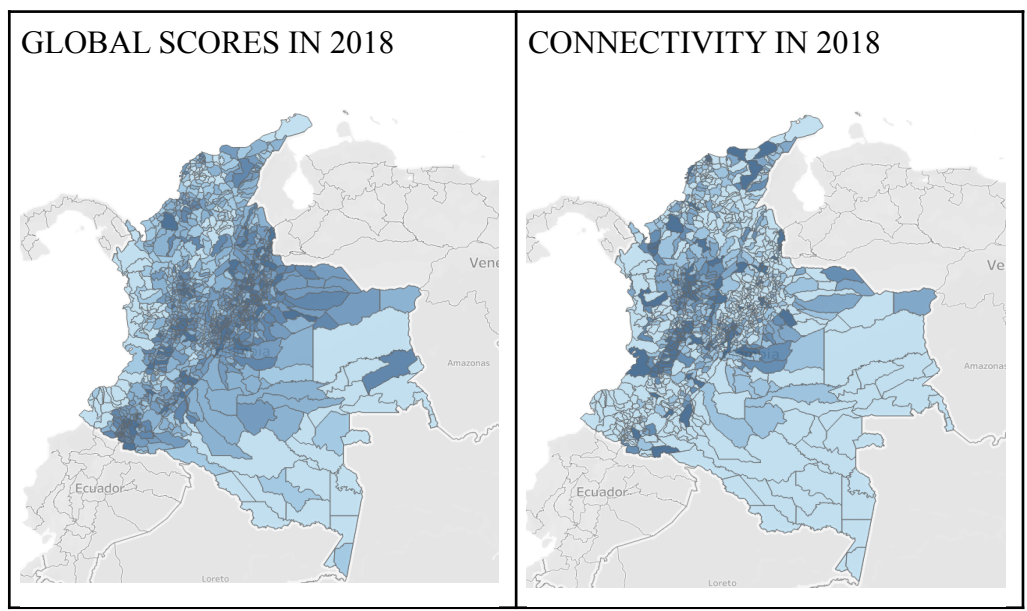

Figure 1. *Global scores and connectivity in 2018 from the aggregated data frame. Dark indicates better score/connectivity.*

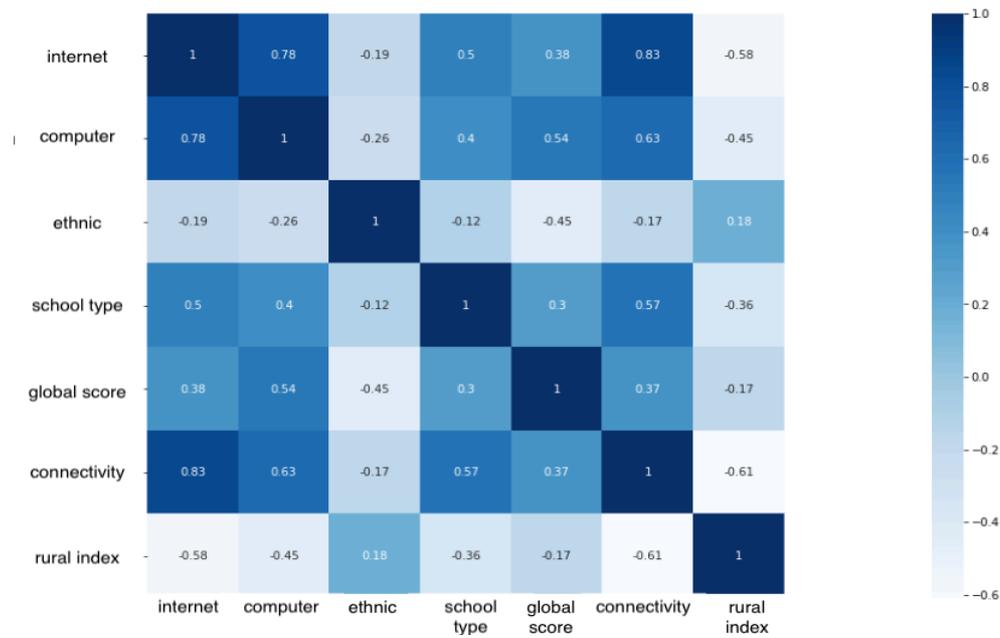

Figure 2. *Correlation heat map for the initial set of variables.*

SCORE, and finally, CONNECTIVITY also has a negative correlation with RURAL INDEX. It is important to highlight that we continue considering COMPUTER/CONNECTIVITY and INTERNET as covariables, even though they are highly correlated since the goal of this work is to give indications/ policies about if it is better to invest on computers, on the internet, or in other issues.

*Classification.* There are many algorithms, from linear and logistic regressions to neural networks, to help identify and classify covariables/results. However, we wanted to choose interpretative models, this is, models that classify but also give us a statistically significant set of covariables. In the literature, it is common to use linear or logistic regression (Bassok & Latham, 2017; Zhao, Lu, Huang & Wang, 2010; Chica, Galvis, Ramírez, 2011; Paulsen and Mccormick, 2020; Rickles, Heppen, Allensworth, Sorensen and Walters, 2018; Riegle-Crumb, King, and Irizarry, 2019) but not common the use of complex algorithms of machine learning (Amorim, Jeon, Abel, Felisberto, Barbosa and Martins, 2020; Kelly, Olney, Donnelly, Nystrand and D'mello, 2018; Xu, Wang, Peng and Wu, 2019; Garcia-Gonzalez, Skryta, 2019). As Hellas et.al (2018) pointed out, outside of classification, there are also clustering, mining and statistical methods in the literature to predict academic performance across the world. Linear modeling is one of the most selected methods, followed by probabilistic graphical models and decision trees.

In particular, in the work of Xu, Wang, Peng, and Wu (2019), they have covariables as online duration, Internet traffic volume, connection frequency, among others. And using decision trees, neural networks, and support vector machines predict academic performance from these features. In this work, we want to do something similar using another set of covariables with logistic regression, decision forest, and regression forest.

Therefore, the initial model is a logistic regression given that this model shows statistically significant covariables and how they affect (positive or negative) the independent variable. Going deeper into the list of models, we found the random forests, which is an ensemble learning method for classification, regression, and other tasks that operates by constructing a multitude of decision trees at training time and outputting the class that is the mode of the classes (classification) or mean prediction (regression) of the individual trees. Random decision forests correct for decision trees' habit of overfitting on the training set.

We construct a classification algorithm over the municipalities considering the initial set of covariables, which gives a list of municipalities at different vulnerability levels and a few actionable features on which governments (national, state, and/or municipal) can act upon. To use these algorithms, we need a risk threshold that we define for each year as the *global score average* minus $k$ times the *global score standard deviation* (where $k$ can be any real number between 0 and 2; for optimal purposes we choose $k$=1). Also, we split the data frame into training data (from 2014 to 2018) and validation data

(2019), to decide about the accuracy. The main results of the logistic regression are:

- Considering the p-value for each covariable (less than 0.05), we conclude that the relevant variables are INTERNET, COMPUTER, ETHNIC, CONNECTIVITY, and RURAL INDEX.
- ETHNIC and INTERNET have positive coefficients which imply these covariables have a positive impact on the global score.
- COMPUTER and RURAL INDEX have negative coefficients which imply these covariables have a negative impact on the global score.

With this new set of covariables, we perform a regression random forest with depth $m$ (with $m>2$; we choose $m=3$) and a classifier random forest of depth $l$. We compare the three models using the AUC, see Table 1 (Area under the curve ROC): The ROC curve is a performance measurement for classification problems at various thresholds settings. ROC is a probability curve and AUC represents a degree or measure of separability. It tells how much a model is capable of distinguishing between classes. The higher the AUC, the better the model is at predicting 0's as 0's and 1's as 1's. The choice of parameters $k, m, l$ was made looking for AUC between 0.7 and 0.87 since lower AUC implies that the model is underfitting and bigger implies that it is overfitting. Notice that logistic regression, according to AUC, is the best classificatory up to these conditions. If we increase the depth to any random forest, we get greater AUC, Table 1.

**AUC for each classification algorithm.**

| MODEL | AUC |
| --- | --- |
| Logistic regression | 0.8684 |
| Regression random forest (depth 3) | 0.8653 |
| Classifier random forest (depth 3) | 0.7032 |

but the complexity of the algorithm increases. Each one predicts over each municipality if it will be at risk or not, that is, if the average global score of the municipality is greater or not than the risk threshold. To compare and use at the same time the results of the three algorithms, we construct a vulnerability level called TOTAL_RISK as the sum of the predicted risk of these three models. This analysis allows to define 4 levels of vulnerability:

1. No vulnerability if TOTAL_RISK=0: none of the models marked it as a municipality at vulnerability,
2. Low vulnerability if TOTAL_RISK=1: one of the models marked it as vulnerable,
3. Medium vulnerability if TOTAL_RISK=2: two of the models marked it as vulnerable,

4. Serious vulnerability if TOTAL_RISK=3: all the models marked it as vulnerable.

In Figure 3, we can see a map of Colombia with all levels of TOTAL_RISK. To determine how good the estimations are, we can compute a confusion matrix, which is a specific table layout that allows visualization of the performance of the algorithm. Each column of the matrix represents the instances in a predicted vulnerability while each row represents the instances in actual risk, and determines if the algorithm is confusing classes. In this case, the first row of the matrix is *(914, 21, 5, 8)* and the second row is *(83, 14, 9, 58)*. For example, the last column *(8, 58)* implies that 66 municipalities are at serious vulnerability (8 of them are false positives, that is, 8 municipalities are labeled as 'at vulnerability' but they are not). We notice that the quantity of false positives and false negatives (83) is little; therefore, the predictions are accurate.

*Beyond classification.* In Figure 3, we also notice some interesting municipalities in the center of the country, in high vulnerability, contradicting the commonly believed that remote locations are always vulnerable. This implies that not necessarily being surrounded by no vulnerability affects completely this covariable.

One of the findings was Coyaima in Tolima, in the center of the country. In Tolima, during the last years, the percentage of the population belonging to an ethnic group is near to 5%, and the percentage of rurality is around 50%, but when we choose Coyaima, those numbers dramatically scale; the percentage of belonging to an ethnic group is over 70% in the last years and rurality is 84%. Reflecting one of the main conclusions: belonging to an ethnic group highly affects vulnerability, controlling access to technologies.

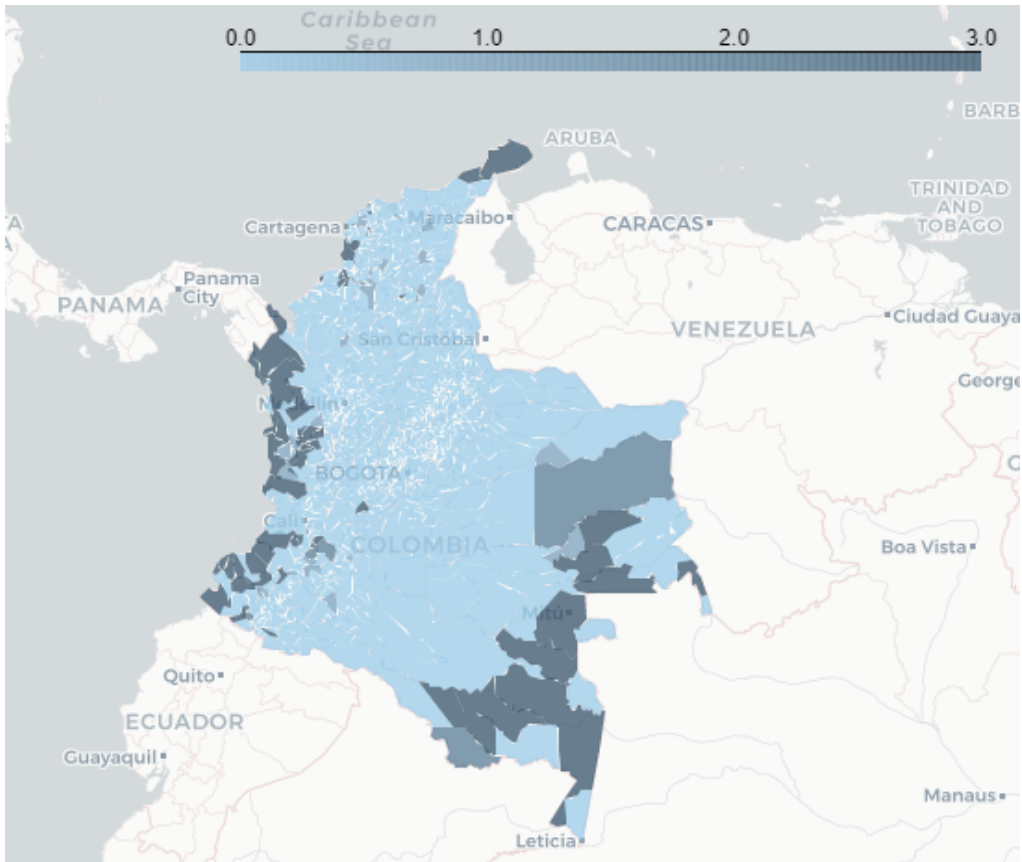

Figure 3. *Vulnerability levels in Colombia. Dark indicates high TOTAL_RISK.*

Changing to Altos del Rosario in Bolivar, in the north of the country, we notice the same pattern: a high percentage of ethnic population and rurality but few internet connections and little computer ownership. However, we can also have some municipalities as Leticia in Amazonas (the last point at the south) without vulnerability. Taking a look at the state as a whole, the percentage of ETHNIC and RURAL INDEX are over 80%, and COMPUTER and INTERNET below to 20%; but in Leticia, we get different values, it is important to note that Leticia is the capital of this state.

We can also evaluate the percentage of municipalities at each vulnerability level by state, this is an interesting point of view because it can be seen that states with a large number of municipalities with high vulnerability have some municipalities with no vulnerability, for example, Choco and Amazonas. These zones can be a reference to improve on their neighbors. Listing states with the count of municipalities in each vulnerability we found that Vaupés, Chocó, Guainía, and Amazonas are states that need serious intervention as soon as possible since almost 67% of its municipalities are in serious vulnerability.

Running a Bonferroni to understand each vulnerability level, we notice that low and no vulnerability have significant differences in self-reported internet, computer ownership, belonging to an ethnic group, and connectivity. Low and medium vulnerability in connectivity; and medium and serious vulnerability in ethnic group and connectivity. Since connectivity and belonging to an ethnic group are common variables, in Table 2, we analyze the average number of connections per 1000 habitants and percentage of students belonging to an ethnic group, in each vulnerability level, obtaining that municipalities at serious vulnerability have a high percentage of students belonging to an ethnic group, opposite to no vulnerability municipalities.

Table 2.

**Average in two covariables according to vulnerability.**

| Vulnerability | Average CONNECTIVITY | Average ETHNIC |
|---|---|---|
| No | 42.92 | 4% |
| Low | 12.94 | 38% |
| Medium | 2.73 | 51% |
| Serious | 7.54 | 89% |

*Improvements*. Next, we can use the vulnerability model to determine if an improvement on the internet and/or computer ownership estimates that the municipality/state is no longer vulnerable. Considering Santa Lucia in Atlántico, in recent years this municipality has been the focus of attention due to a flood when the dam channel broke. Because of this, there are many social initiatives in this municipality. In 2019, was reported a percentage of 6.7% on INTERNET, 24.6 % on COMPUTERS, 77.5% on ETHNIC, CONNECTIVITY of 15%, RURAL INDEX of 14%, and an average global score of 201.26. According to this model, this municipality is highly vulnerable. With this model, if we improve 23% on computers we get low vulnerability, showing how better conditions in the house to study (such that, having a computer) could improve academic performance.

Now, consider a State as Amazonas, in the south of the country. As we notice before, more than 70% of its municipalities are in serious vulnerability. Doing a 47% improvement on computer ownership we get medium vulnerability, however, no other improvement gets low or no vulnerability. Adding 144 internet connections vulnerability changes to medium. Then state policies also could reflect better academic performance.

RESULTS.

With COVID isolation measures banning students from classrooms, most learning has moved online, but even those pupils who can successfully connect are likely to fall behind if not possessed of the self-direction and motivation needed for remote education. Moreover, research has shown that poorer students perform worse in online courses (Chetty, Friedman, Hendren & Stepner, 2020) than face-to-face ones.

Our analysis found that the most relevant covariables related to an increased academic vulnerability in Colombia, related to technologies, are Connectivity per 1000 inhabitants (as a measure of reliable broadband internet access) and belonging to an ethnic group (associated with low income). Amazonas, Vaupés, Guainía, and Chocó need serious intervention, given the high values shown for the vulnerability factors measured. Geographic centralization (associated with increased urbanization, as measured by the Rural Index), is also a strong predictor of better scores for students.

We found some centralized municipalities with high vulnerability as Coyaima in Tolima and Altos del Rosario in Bolívar, surrounded by municipalities of zero vulnerability. In particular, these municipalities have low access to the internet and computer and high rurality and percentage of belonging to an ethnic group, showing an interesting feature of Colombia.

According to Hartley and Bendixen (2010), improving the academic results and reducing the large gap in academic achievement requires the interventions to focus on maintaining the students' interactions (among themselves and with their teachers), also learning skills in the students are necessary as self-regulatory skills, epistemological beliefs, motivation, self-efficacy, ability, physical challenges, and learning disabilities. For this scenario to be realized, a series of factors have to line up: schools need to have the resources to implement remote learning, students need to have access to computers and reliable internet connections, and parents need to have the ability, time, energy, and patience to turn into home-school instructors.

A concerted effort between the state government and parents seems to be the most effective strategy. Complementary projects might include:

- Optimizing accessible solutions to mobile devices given that students have some access to them.
- Offering limited data plans (access granted only to academic sites, to avoid misuse).
- Focused policies on the rural ethnic minorities, the communities with higher vulnerability.
- Lending computers to families.
- Delivering financial stimulus to the municipalities most at vulnerability, conditional on the improvement of academic results compared with the previous year.
- Support to parents so they can step up as temporary teachers.

The challenges are related to providing students at vulnerability the necessary tools for success, not only laptops and reliable broadband internet access but also to motivate parents to help their children succeed academically.

NOTES. Interested readers can refer to the original data set to see how covariables were precisely coded https://github.com/team63/remote_education.git. Also, at https://public.tableau.com/profile/sergio.daniel.segura#!/vizhome/shared/69CQ6C95X and https://

[public.tableau.com/profile/sergio.daniel.segura#!/vizhome/Descrip_Modelo/Departamento](public.tableau.com/profile/sergio.daniel.segura#!/vizhome/Descrip_Modelo/Departamento), the authors developed two interactive dashboards on Tableau public. The first one, with statistics by municipalities, states, and years over the covariables GLOBAL SCORE, CONNECTIVITY, RURAL INDEX, ETHNIC, INTERNET, and COMPUTER; the other one we add the new variable TOTAL_RISK.

REFERENCES.

AUTHORS.

ADRIANA MEJIA CASTAÑO, PhD, is an adjoint professor of mathematics and statistics at Universidad del Norte, km 5 via Puerto Colombia, Barranquilla - Colombia, mejiala@uninorte.edu.co, +57 3058134462. Corresponding author.

JAVIER E. HERNANDEZ, Mg. Management, Independent Researcher, Cra 15 #110 18, Bogotá - Colombia, jeheca85@yahoo.com, +57 3229488884.

ANGIE MENDEZ LLANOS, Mathematician, Independent Researcher , Calle 128 F # 104 05, Bogotá - Colombia, almendezl@unal.edu.co, +57 3213474300.

ACKNOWLEDGEMENTS. This work was realized under the support of Correlation One, MINTIC, and the program DS4A, the second version in Colombia. We thank the other members of team 63: DIEGO AVILA, DIEGO SEGURA, JUAN CARREÑO, and SERGIO SEGURA.

APPENDIX

Here we collect descriptive statistics over the aggregated data frame. Analyzing the main covariables across time (COMPUTER, INTERNET, ETHNIC and RURAL INDEX), we notice that the percentage of students belonging to an ethnic group, access to computers, and rurality are almost the same in recent years, however, access to the internet appears to grow, from a global coverage of 20% to 33%  (Figure I-a). Analyzing only the main states (Cundinamarca, Antioquia, Valle del Cauca, Atlántico and Magdalena, see Figure I-b) we notice a similar behavior but comparing with Figure I-a, the percentage belonging to an ethnic group and rurality decrease (from 11% to 3% and from 56% to 51%, respectively), access to computer and internet increase; showing, as we expected, large differences in these demographic aspects. Analyzing the main municipalities (Bogotá, Medellín, Cali, Barranquilla, and Cartagena), the variations are remarkable (Figure I-c) over COMPUTER and INTERNET ACCESS (highly increased) and RURAL INDEX (almost disappeared).

Focusing on the global score in the main states in 2019 we notice that good scores (in dark blue) are usually near to the capital (Figure II), showing that investment in quality education is performed mainly in big cities, except in Bolivar (capital Cartagena) where the academic performance is low. However, Cundinamarca has good scores in almost every municipality, we assume that this happens because Bogotá (the country capital) is in this State.

We now analyze the levels of vulnerability predicted by our algorithm, remember that the choice of the threshold was made to detect the municipalities most at risk. We found that 90% of municipalities in the country obtain no risk. This could be an indicator that, for future works, the threshold risk could be the average in every year. At low risk is 1%, in medium risk 3%, and in high risk 6%.

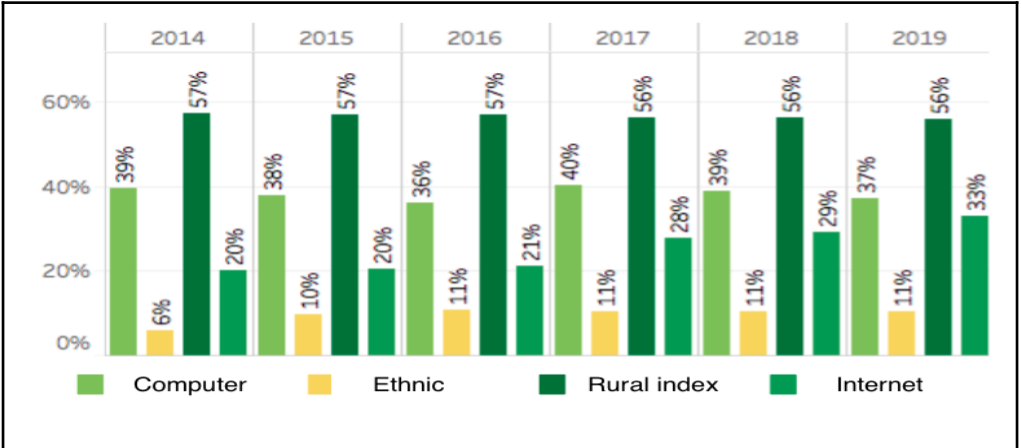

Figure I-a. *Main covariables across time in the entire country.*

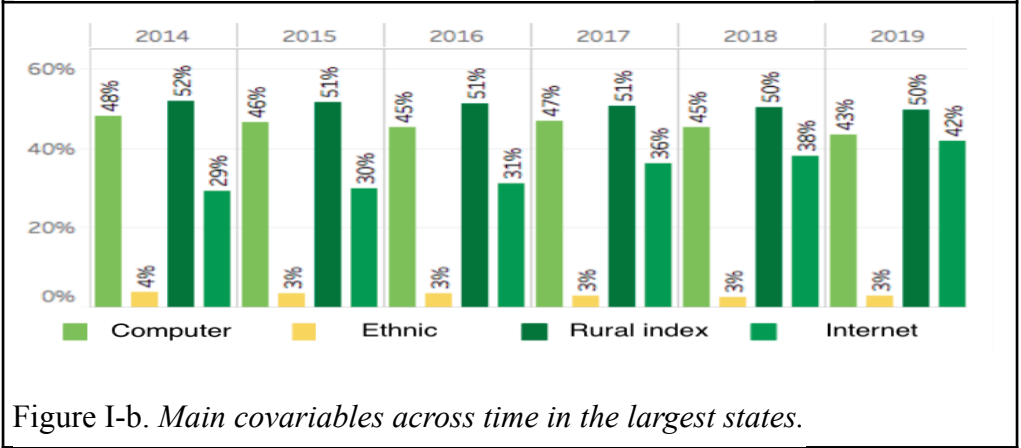

Figure I-b. *Main covariables across time in the largest states.*

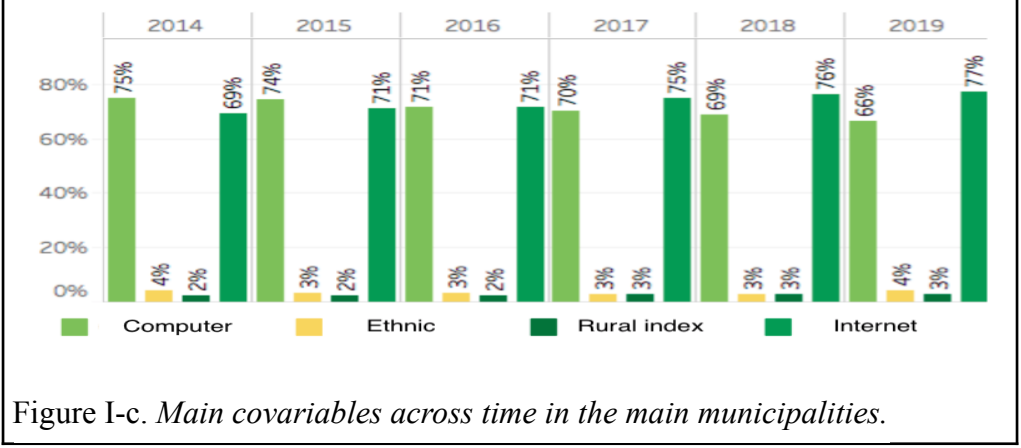

Figure I-c. *Main covariables across time in the main municipalities.*

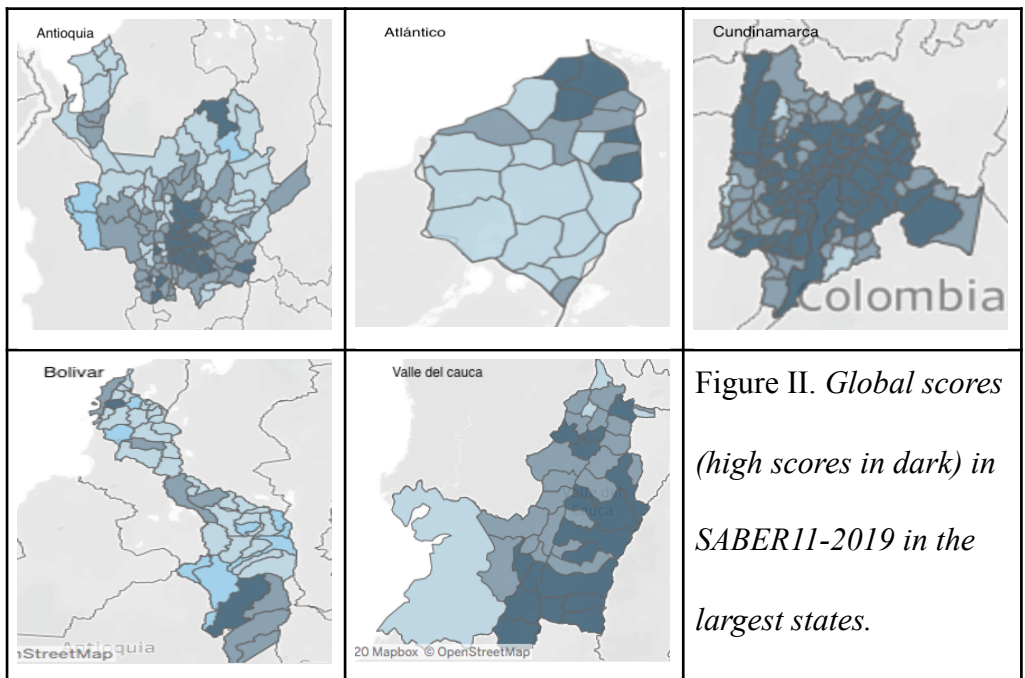

Figure II. *Global scores (high scores in dark) in SABER11-2019 in the largest states.*

As mentioned before, the municipalities in high vulnerability are mainly in remote locations, the percentage of computers across time is stable, the percentage of internet access has increased since 2017, but the percentage of students belonging to an ethnic group and rurality are high (near 70%). In medium vulnerability municipalities, ETHNIC is around 52% and RURAL INDEX 67%, showing that belonging to an ethnic group is decisive for this classification. In low vulnerability, ETHNIC values are around 39% and RURAL INDEX 50%; and in no vulnerability, the values are 4% for ETHNIC and 55% for RURALITY. Showing that at this level ETHNIC is too decisive. Finally, we want to analyze the relation of connectivity (access to internet reported by MINTIC), RURAL INDEX, GLOBAL SCORE, and TOTAL_RISK (Figure III): The municipalities in high risk (dark blue) have poor connectivity, showing that it is really necessary to improve internet connection across the country as a technology measure to improve academic performance.

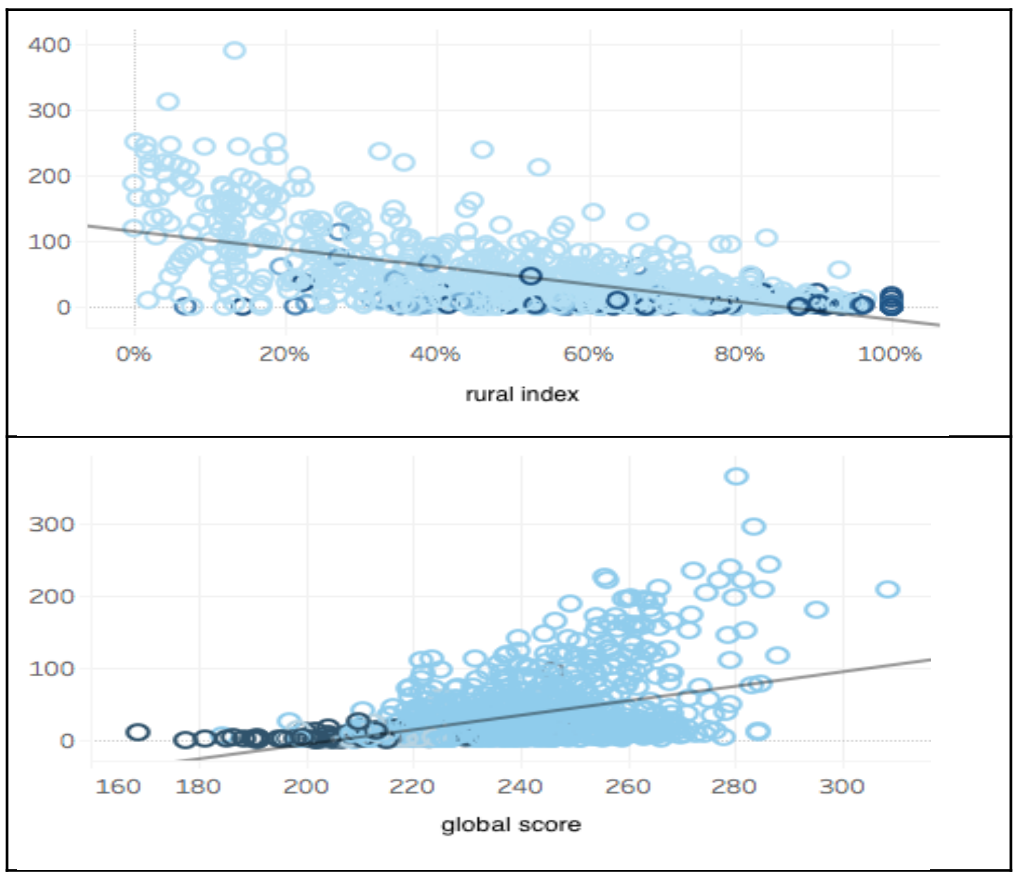

Figure III. *Comparison between RURAL INDEX, GLOBAL SCORE, CONNECTIVITY (y-axis), AND TOTAL_RISK. Dark circles show high vulnerability.*